\documentclass[twoside]{article}

\usepackage[sc]{mathpazo} 
\usepackage[T1]{fontenc} 
\usepackage{microtype} 

\usepackage[hmarginratio=1:1,top=32mm,columnsep=20pt]{geometry} 
\usepackage[hang, small,labelfont=bf,up,textfont=it,up]{caption} 
\usepackage{booktabs} 
\usepackage{float} 
\usepackage{hyperref} 

\usepackage{lettrine} 
\usepackage{paralist} 

\usepackage{abstract} 

\usepackage{titlesec} 
\renewcommand\thesection{\Roman{section}}
\titleformat{\section}[block]{\large\scshape\centering}{\thesection.}{1em}{} 

\usepackage{fancyhdr} 
\title{\vspace{-15mm}\fontsize{24pt}{10pt}\selectfont\textbf{Information-Theoretic Security\newline for the Masses}} 

\author{
\large
\textsc{Oleksandr Nikitin}\footnote{This is my first TeX-formatted article, and first contribution to the cryptography field, any feedback is greatly appreciated}\\[2mm] 
\normalsize https://tvori.info \\ 
\normalsize \href{mailto:oleksandr@tvori.info}{oleksandr@tvori.info} 
\vspace{-5mm}
}
\date{}

\begin{document}
 
\maketitle 

\thispagestyle{fancy} 


\begin{abstract}

We combine interactive zero-knowledge protocols and weak physical layer randomness properties to construct a protocol which allows bootstrapping an IT-secure and PF-secure channel from a memorizable shared secret. The protocol also tolerates failures of its components, still preserving most of its security properties, which makes it accessible to regular users.

\end{abstract}



\section{Introduction}

\lettrine[nindent=0em,lines=2]{E} xisting session key derivation schemes are problematic. We want to achieve perfect forward secrecy and information-theoretic security, while imposing minimal burden on end users and assuming they will make errors from time to time, such as losing devices with sensitive data or ignoring MITM warnings.

\section{Changelog}

\begin{compactitem}
\item \textbf{Version 0.3}: Minor corrections, arXiv.org release.
\item \textbf{Version 0.2}: Clarified\footnote{Thanks to sassa\_nf for raising this question} the relationship of RSA security, DH security and recommended lifetime of X (the only long-lived value).
\item \textbf{Version 0.1}: Initial public release.
\end{compactitem}


\section{Initial Conditions}

Participating Parties:

\begin{compactitem}
\item Alice, who has some knowledge previously shared with Bob (which, however, is not enough by itself to provide enough entropy to use as a shared secret), called X.
\item Bob, sharing the same unit of knowledge with Alice
\item Mallory, who for some reasons wants to listen to conversation between Bob and Alice and can record each and every packet passing in-between.
\end{compactitem}

We also assume that both Alice and Bob have access to a cryptographically secure RNG.

\section{Mutual Authentication}

\begin{enumerate}
\item Alice and Bob each generate an one-time, ephemeral RSA key pairs KA1, KB1.
\item They exchange the public keys for KA1 and KB1. From this point, they can send encrypted messages but cannot be sure of each others' identity.
\item To authenticate each other, Alice and Bob execute a SMP protocol, as described in [SMP07], verifying that their views of the system (fingerprints of the public keys and the value of X) are indeed the same. From this point, the channel can be considered computationally-secure, but provides neither PFS nor repudiability. Since the secret X may be long-lived, it makes sense to strengthen X, e.g. make the parties agree on H(KA1pub, KA2pub, PBKDF2(X)), and/or agree on a random salt for X beforehand.
\item Alice generates a random number RA1, while Bob generates RB1. They exchange these numbers, generating the one-time shared secret RS1 as H(RA1,RB1) where H is a cryptographic hash function.
\end{enumerate}


\section{Securing}

\begin{enumerate}
\item Alice and Bob run a Kish-Sethuraman protocol [KS04] using packet timing variation as an entropy source, as described in [ITP13] to produce the shared secret RS2.
\item Secrets RS1 and RS2 are combined to produce the shared symmetric key RS3.
\item Alice and Bob exchange KA1 and KB1.
\item KA1, KB1, RA1, RB1, RS1, RS2 are erased.
\item Shared key RS3 is used for subsequent communication of data messages. If the key RS2 was not derived by Mallory at this point, breaking KA1 and KB1 is no longer feasible.
\end{enumerate}

\section{Termination}

\begin{enumerate}
\item (optional) Alice and Bob establish new shared secret X2.
\item Alice and Bob erase RS3.
\end{enumerate}


\section{Attacks}

As should be evident, the resulting key RS3 is composed from the information derived from the packet timing information, and the entropy from Alice and Bob RNGs.

\textbf{Scenario 1}. With access to DPI and packet forwarding hardware with precise packet timestamping facility, and the value of X, Mallory mounts a MITM attack, computes the shared secret RS1, and obtains precise packet information enough to extract the same bits as Alice and Bob do.

\textbf{Result}: Complete protocol failure. However, no information is captured that allows decrypting previous or future sessions.

\textbf{Scenario 2}. With the ability to do precise packet timestamping, but without knowing X, Mallory records the conversation trace.

\textbf{Result}: The security of the protocol is reduced to that of RSA (since knowledge of KA1 and KB1 allows Mallory to partially deduce RS1, and precise packet timing allows to deduce RS2).

\textbf{Scenario 3}. Mallory has access to packet mirroring hardware (typical to the interception facilities deployed today)

\textbf{Result}: After breaking KA1 and KB1 the integrity is preserved. When the hardware can also record precise timestamps for each packet, then the information-theoretic security decreases accordingly. However, unless the Mallory can measure timings directly at both Alice and Bob's computer, the amount of information extracted will not be enough to completely derive RS2, as shown in [ITP13].

\textbf{Scenario 4}. Mallory interrogates Bob to leak session plaintext and/or X, or Bob leaks this data unintentionally.

\textbf{Result}: Mallory receives the session plaintext and the secret X. This allows Mallory to mount a MITM attack, as in Scenario 1, unless Alice and Bob agree on a new X in the meantime. Other security properties are not compromised, since sessions are completely unrelated to each other.

\textbf{Scenario 5}. Mallory is able to break both RSA, DH and chosen PBKDF in reasonable time (e.g. hours or days)

\textbf{Result}: Mallory recovers X, which can be used to easily break into subsequent sessions. So, the X must be renewed over the ITS channel from time to time, if such an attacker is to be expected.


\section{Benefits}

\textbf{IT-secure perfect forward secrecy}: Unless the information sufficient to derive RS2 was captured during the communication session, it will not be possible to decrypt the session trace given any amount of computing power. In addition, no information derived from breaking one session will help breaking another.

\textbf{Graceful degradation}: Even when the attacker has access to exactly same timestamp information, the security of the protocol is at least as good as RSA, because of the construction of key RS1.

\textbf{Repudiation}: Since all keys except X are entropy-based, it is impossible to correlate any two communication sessions between the same parties based on the protocol information alone. (Of course, after the protocol failure, data being exchanged will probably allow correlation, but IT-PFS means an attacker would have to break all sessions to do that)

\textbf{Deniability}: Since no user-originated messages are encrypted using KA1 or KB1, both Alice and Bob have seen KA1 and KB1 at the end of "Securing" stage, and data messages are not signed in any way, neither Bob nor Alice can use protocol information alone to prove his counterparty said something.

\textbf{Accessibility}: Alice and Bob do not need to care about secure key distribution. Leaking X post-factum does not help Mallory with breaking previous traces, either. The X can be simple (e.g. 10-bit secure), and therefore can be memorized.

\section{Drawbacks}

\textbf{Session setup time} As shown in [KS04], time required to derive RS2 is around 20-30 minutes. However, this key can be long-lived, and periodically renewed in the background without further user intervention. 

During the initial RS2 generation period, Alice and Bob can fall back to e.g. OTR, while being informed via application UI that the channel is not yet fully secured.

\section{Conclusion, Discussion}

This article presents an easily-accessible way for users to bootstrap the memorizable shared secret into an IT-secure channel in the presence of an attacker with (virtually) infinite computing resources (the RSA keys must only withstand bruteforce for the duration of Kish-Sethuraman protocol).

Still, if Mallory does really have unlimited computing resources, once you successfully established the RS3 once, you can't rely on SMP or RSA for next sessions, of course.

But if the resources of Mallory are limited, albeit large, then the security of the protocol boils down to the ability of renewing X faster than Mallory can break RSA, DH and PBKDF to recover X from the session traces.

Leaking or bruteforcing the secret does not compromise previous sessions based on it. Also, even when the secret is known to the attacker, he still needs to mount a full MITM for each future session. 

The key RS2 can be generated as H(RA1,RB2,X), but that does not really affect security properties, since knowing RA1/RB2 we can still bruteforce RS2 (but we can't bruteforce X when the SMP authentication is in progress)


\end{document}